\documentclass[showpacs,12pt]{revtex4}
\usepackage{epsfig}
\usepackage{epsf}
\usepackage{color}
\usepackage{graphicx}
\usepackage{amsmath}
\def\beq{\begin{equation}}
\def\eeq{\end{equation}}
\def\bea{\begin{eqnarray}}
\def\eea{\end{eqnarray}}
\begin{document}

\title{Compressibility of rotating black holes}
\author{Brian P. Dolan}
\email{bdolan@thphys.nuim.ie}
\affiliation{{Department of Mathematical Physics, National University
of Ireland,}\\
{Maynooth, Ireland}\\
{\rm and}\\
{Dublin Institute for Advanced Studies,
10 Burlington Rd., Dublin, Ireland}\\
}

\begin{abstract}

Interpreting the cosmological constant as a pressure, whose
thermodynamically conjugate variable is a volume, modifies the
first law of black hole thermodynamics by adding a $PdV$ term,
giving a complete form of the first law.
Properties of the
thermodynamic volume are investigated: the compressibility and the
speed of sound of the black hole are derived in the case of non-positive
cosmological constant. The adiabatic compressibility 
vanishes for a non-rotating black hole
and is maximal in the extremal case --- comparable with, but still less than,
that of a cold neutron star.  A speed of sound
$v_s$ is associated with the adiabatic compressibility, which is 
is equal to $c$ for a non-rotating black hole and decreases as the angular momentum is increased.  An extremal black hole
has $v_s^2=0.9 \,c^2$ when the cosmological constant vanishes,
and more generally $v_s$ is bounded below by $c/ {\sqrt 2}$. 

\end{abstract}


\maketitle
\section{Introduction}

The first law of black hole thermodynamics has had a long and subtle
history, starting with Bekenstein's proposal \cite{Bekenstein}
that the entropy should be proportional to the area followed by 
Hawking's seminal work on black hole-temperature \cite{Hawking}. 
Notably absent in the discussion of the first law
has been a $PdV$ term, associated with changes in volume.
It is shown here that, for rotating black holes, a thermodynamic volume
can be defined which is independent of the area allowing a complete form
of the first law to be formulated.
 
It has been suggested by a number of authors
\cite{HT}-\cite{CCK}
that, for a black hole of mass $M$ when a cosmological constant is included,
the first law of thermodynamics should read
\beq \label{firstlaw}
d M = TdS + \Omega d J + \Phi d Q + \Theta d\Lambda,
\eeq
where $S$, $J$ and $Q$ are the entropy, angular momentum
and charge respectively while $\Theta$ is a thermodynamic variable conjugate to the cosmological constant $\Lambda$.
Since $\Lambda$ is associated with a pressure, $\Lambda = - 8\pi P$
in units with $G=c=1$, one interpretation of the $\Theta \,d\Lambda$ term
is that it is related to the familiar term $PdV$ term in the first law, 
with $\Theta$ proportional to the volume.  This 
in turn requires interpreting
the black hole mass as corresponding to the thermodynamic
enthalpy, $H=M(S,P,J,Q)$, rather than the more usual interpretation as the thermal energy $U$, \cite{KRT,Enthalpy}.  One then has
\beq \label{dH}
d M = TdS + \Omega d J + \Phi d Q + VdP, \eeq
where the thermodynamic volume, 
\beq
V=\label{Vdef}
\left.\frac{\partial M}{\partial P}\right|_{S,J,Q}=-8\pi \Theta,\eeq 
is the Legendre transform of the pressure.  In full generality, the first law of black hole thermodynamics now reads
\beq \label{FirstLaw}
d U = TdS + \Omega d J + \Phi d Q -PdV, \eeq
where $U=H- P V$.   In this work
we shall set $Q=0$ for simplicity, our
results are easily extended to include the $Q\ne 0$ case.

Of course if $\Lambda$ is truly constant, $P$ cannot be varied and the last term in (\ref{dH}) is always zero, though the last term in (\ref{FirstLaw}) need not be zero and can have physical consequences \cite{PdV}.
But it is commonly supposed, in inflationary models for example,
that $\Lambda$ has at some time in the past been significantly different from its current value and so must vary. 

Unfortunately the thermodynamics of black holes with $\Lambda>0$ are not well understood: such a system has two different temperatures in
general, one associated with the black hole horizon and one with the cosmological horizon, and in addition there is no asymptotic regime in which an invariant mass can be defined.
For these reasons we shall primarily restrict ourselves to the better
understood $\Lambda<0$ case in this work.  This is
sufficient to access the $\Lambda=0$ limit.

The mass of a charged AdS-Kerr black hole, with cosmological constant $\Lambda=-8\pi P$, was expressed as a function of $S$, $J$ and $Q$ in \cite{CCK}.  When $Q=0$
the expression is
\beq \label{CCKmass}
M(S,P,J)=
\frac {1}{2}\sqrt{\frac{\left( 1+\frac{8 P S}{3}  \right)
\left\{S^2\left(1 + \frac{8 P S}{3}\right) + 
4 \pi^2 J^2\right\}}
{\pi S}}\;,
\eeq
where $G=c=1$ and $S$ is $\frac 1 4$ the area of the event horizon.
From this the temperature can be evaluated,
\beq \label{CCKTemperature}
T= \left.\frac {\partial M}{\partial S}\right|_{P,J}=\frac {1}{8\pi M}\left\{
\left(1 +\frac {8 P S}{3} \right)
\left(1 + 8 P S \right)
-4\pi^2 \frac {J^2}{S^2}\right\},
\eeq
with $\hbar=1$.
The temperature vanishes for extremal black holes, with 
maximum angular momentum
\beq \label{Jextremal}
J^2_{max}=
\left(\frac {S}{2\pi}\right)^2\left(1 +\frac {8 P S}{3} \right)
\left(1 + 8 P S \right).
\eeq

The stability properties of Kerr-AdS black holes were studied in \cite{CCK},
where it was shown that there is a critical point near $J=0.0236$ (the authors of \cite{CCK} used units in which $P=\frac{3}{8\pi}$).

The thermodynamic volume (\ref{Vdef}),
\beq\label{ThermodynamicVolume}
V= \left.\frac {\partial M}{\partial P}\right|_{S,J}=
\frac{2}{3 \pi M}\left\{S^2\left(1 + \frac{8 P S}{3}\right)+ 2\pi^2 J^2 \right\},
\eeq
is always greater than the na{\"\i}ve geometric result \hfill\break
$V=\frac{4\pi}{3}\left(\frac{S}{\pi}\right)^{\frac 3 2}$, 
as observed in \cite{CGKP}.

The Legendre transform of (\ref{CCKmass}),
\beq \label{InternalEnergy}
U(S,V,J) =
\left(\frac {\pi}{S}\right)^3\left\{ 
\left(\frac{3V}{4\pi}\right)
\left( \frac {S^2} {2\pi^2} +J^2 \right) 
-J^2\sqrt{\left(\frac{3V}{4\pi}\right)^2-\left(\frac S \pi \right)^3}
 \right\},
\eeq
was studied in \cite{PdV},
where the difference between the specific heat at constant pressure
\beq C_P=T\left.\frac{\partial S}{\partial T}\right|_{P,J}
=T\frac {1}{\left.\frac{\partial^2 H}{\partial S^2}\right|_{P,J}}
\eeq
and specific heat at constant volume
\beq C_V=T\left.\frac{\partial S}{\partial T}\right|_{V,J}
=T\frac {1}{\left.\frac{\partial^2 U}{\partial S^2}\right|_{V,J}}
\eeq
was also explored. 
Explicit expressions confirm that   $\frac{C_P}{C_V}\ge 1$.
$C_P$ diverges along the curve in the $J-S$ plane shown in figure 1 (plotted using dimensionless variables $JP$ and $SP$).  $C_P$ is negative in the region below this curve and the critical point corresponds to the maximum value of $JP$ which is determined numerically to lie at $JP=0.00286\ldots$ and $SP=0.0820\ldots$\;. 
When $J=0$,
\beq
C_P=\frac{2 S (1+8PS)}{8PS-1}
\eeq
which is negative when $P=0$, the famous negative
heat capacity of Schwarzschild black holes refers to
$C_P$. 

This instability is reflected in the isothermal compressibility 
\bea
\beta_T&=&-\frac {1} {V} \left( 
\frac{\partial V}{\partial P}\right)_{T,J}\\
&=&\,{\frac {48\,S \left\{ 9\,{j}^{6}+3\, \left( 5+8\,p \right)  \left( 3+8
\,p \right) {j}^{4}+6\, \left( 3+8\,p \right) ^{3}{j}^{2}+ \left( 3+8
\,p \right) ^{4} \right\} }{ \left( 6+16\,p+3\,{j}^{2} \right)  \left\{ 
9\, \left( 9+32\,p \right) {j}^{4}+6\, \left( 3+16\,p \right)  \left( 
3+8\,p \right) ^{2}{j}^{2}+ \left( 8\,p-1 \right)  \left( 3+8\,p
 \right) ^{3} \right\} }},
\nonumber
\eea
where $j=2\pi J/S$ and $p=PS$.
$\beta_T$ diverges when the denominator vanishes,
which is the same locus of points on which $C_P$ diverges and is shown below
in the $JP-SP$ plane appropriate for constant $P$ processes,

\centerline{\vtop{\hbox{\includegraphics[width=200pt,height=200pt]{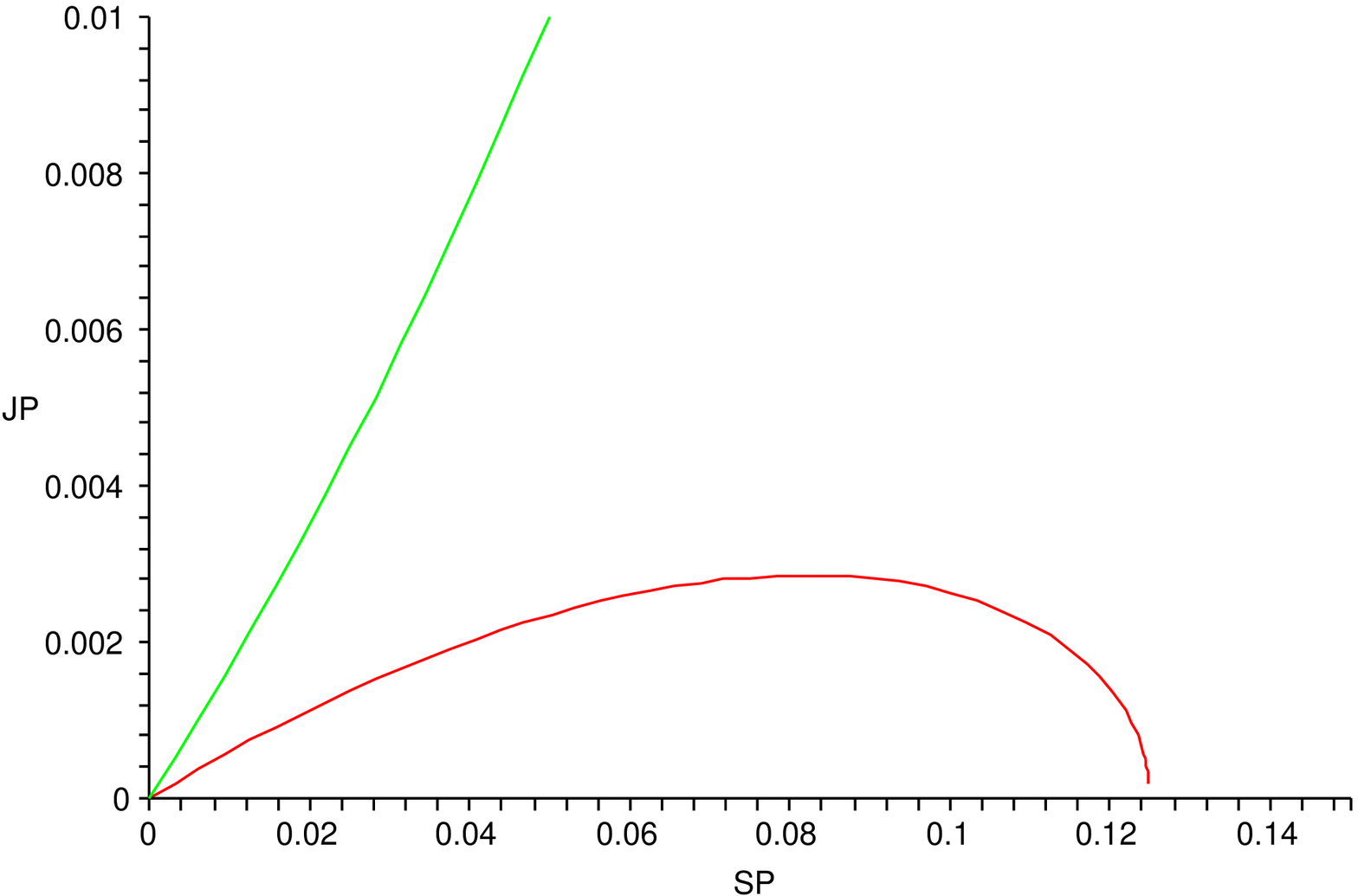}}
\hbox{Figure 1: region in the $JP-SP$ plane in which $C_P$ is} 
\hbox{negative. $C_P$ diverges on the lower curve and is negat-}
\hbox{ive below it.  The upper curve shows $J_{max}$, {\it i.e.} $T=0$,}
\hbox{on which $C_P$ vanishes.}}}
\bigskip


For a non-rotating black hole
\beq \left.\beta_T\right|_{J=0}=\frac{24S}{8PS -1}\eeq
is negative for $PS<\frac 1 8$.

Perhaps a more useful quantity to consider is the adiabatic compressibility,
\beq 
\beta_S=-\frac {1} {V} \left( 
\frac{\partial V}{\partial P}\right)_{S,J}.
\eeq
Evaluating $\beta_S$ from (\ref{ThermodynamicVolume}) shows that it
is manifestly positive and regular,

\beq
\beta_S=
\frac {36\,S j^4}
{\left( 3+{8\,p} \right)  \left(3+ 8\,p + 3\,j^2 \right)
\left( 6+16\,p +3\,j^2 \right)}\;.
\eeq

Thus increasing the pressure, keeping the area and angular momentum
constant, causes the volume of the black hole to decrease, while at 
the same time the mass must increase according to (\ref{CCKmass}). 
$\beta_S$  vanishes when $J=0$, so non-rotating black holes are adiabatically incompressible, but it increases
with $J$ attaining a maximum value in the extremal case, (\ref{Jextremal}), when
\beq\label{Extremalbeta}
\left.\beta_S\right|_{T=0}=
\frac {2\,S \left( 1+8\,p \right) ^{2}}
{\left( 3+8\,p\right)^2  \left( 1+4\,p\right)}.
\eeq

A speed of sound, $v_s$, can also be associated with the black hole, in the usual
thermodynamic sense that
\beq v_s^{-2} = \left.\frac  {\partial \rho}{\partial P}\right|_{S,J}= 1 + \rho \,\beta_S 
=1+
\frac {9\,j^4}
{ \left( 6 + 16\,p + 3\,j^2 \right)^2}\;,
\eeq
where $\rho=\frac M V$ is a density.
This is not the sound associated with any kind of surface wave on
the event horizon, an impossibility due to the no hair theorem,
but rather it is associated with a breathing mode for the black hole
as the volume changes with the pressure, keeping the area constant.
Clearly $0\le v_s^2\le 1$ (this is also true when $Q\ne 0$).
The speed of sound is unity for incompressible non-rotating black holes
and is lowest for the extremal case
\beq \left.v_s^{-2}\right|_{T=0}=
1+\left(\frac{1+8\,p}{3+8\,p}\right)^2,
\eeq
giving $v_s^2=0.9$ when $P=0$ and $v_s^2$ 
achieving a minimum value of $1/2$ when $PS\rightarrow \infty$.

These results show that the equation of state is very stiff 
for adiabatic variations of  non-rotating black holes and softens as $J$ increases.  For comparison, the adiabatic compressibility of a degenerate gas of $N$ relativistic 
neutrons in a volume $V$ at zero temperature
follows from the degeneracy pressure
\beq
P_{deg}=(3\pi^2)^{\frac 1 3}\frac{c\hbar}{4}\left(\frac{V}{N}\right)^{-\frac 4 3 }\quad \Rightarrow \quad \beta_S = \frac 3 {4P_{deg}}.\eeq
For a neutron star $\frac N V\approx 10^{45}\ m^{-3}$ and 
$\beta_S\approx 10^{-34}\ kg \,m \,s^{-2}$.
With zero cosmological
constant the black hole adiabatic compressibility 
at zero temperature is given by (\ref{Extremalbeta}) with $P=0$,
\beq \label{betaZeroP}
\left.\beta_S\right|_{T=P=0}= \frac {2S}{9}=\frac{4 \pi M^2 G^3}{9 \,c^8},\eeq
where the relevant factors of $c$ and $G$ are included.
Putting in the numbers
\beq\left.\beta_S\right|_{T=0}=2.6\times 10^{-38}\left(\frac{M}{M_\odot}\right)^2\ m\,s^2\,kg^{-1},
\eeq
which is still two orders of magnitude less than that of
a neutron star even for $M\approx 10 M_\odot$.  We conclude that the 
zero temperature black hole equation of state is stiffer than that of
a neutron star.

 A subtlety with this analysis of adiabatic compressibility is that, if a black hole is extremal and the pressure goes down, keeping the entropy and the angular momentum fixed, then the extremal value of the angular momentum 
(\ref{Jextremal})
goes down and the black hole ends up with an angular momentum above the new extremal value, giving a naked singularity.  One interpretation of this would be that cosmic censorship
simply does not allow such a process to occur
and, at zero $T$, the entropy must increase, 
if the pressure is decreased, in such a way as to avoid the extremal angular
momentum going down.  If the extremal angular momentum is kept
constant then, from (\ref{Jextremal}), 
\beq
\left.\frac{\partial S}{\partial P}\right|_{T=0}
=-\frac{16\,S^2(1+ 4\,p)}{3+48\,p+128\,p^2}\;.
\eeq
With this constraint the zero temperature compressibility is increased above the
adiabatic value, in fact it becomes
identical to the zero temperature isothermal compressibility,
\beq
\beta_{T=0}={\frac {2\,S \left(11+80\,p+ 128\,p^{2} \right) }{ \left(1+  4\,p \right)  \left( 3+ 48\,p +128\,p^2\right) }}\;.
\eeq
At zero pressure this is a factor of 33 larger than (\ref{betaZeroP}), hence
cosmic censorship effectively makes the zero temperature equation
of state softer.
 
Although the analysis presented here is strictly only valid for $P\ge 0$ nothing obviously goes wrong for $P<0$, provided $PS\ge -\frac 1 8$, and it may be reasonable to analytically continue these thermodynamic relations to negative pressures, \cite{LM}, provided the resulting positive cosmological constant is not too large. 

It has been suggested that mini black holes could have been created in the early universe, \cite{PBH}. 
If this is the case, and they are created with a range of angular momenta, it is interesting to speculate that a collection of such compressible
black holes might affect the speed of sound through the surrounding medium, 
in the same way that a suspension of compressible spheres would affect the speed of sound in a fluid. 

In summary it has been argued that a $PdV$ term should be included
to complete the first law of black hole thermodynamics since
the thermodynamic volume is independent of the area for rotating
black holes.  The famous instability of black holes due to negative heat
capacity is associated with the heat capacity at constant pressure rather
than the heat capacity at constant volume and compressibilities,
together with the speed of sound, have been calculated.


\end{document}